\documentclass[pra,aps,amsmath,amssymb,twocolumn,showpacs]{revtex4}
\usepackage{graphicx}

\newcounter{abcd}

\begin{document}
\title{Noise-induced clustering in Hamiltonian systems}
\author{D.V. Makarov, M.Yu.~Uleysky, M.V~Budyansky, and S.V.~Prants}
\affiliation{Laboratory of Nonlinear Dynamical Systems,\\
V.I. Il'ichev Pacific Oceanological Institute of the Russian Academy of
Sciences,\\
690041 Vladivostok, Russia}
\date{\today}
\begin{abstract}
The motion of oscillatory-like nonlinear Hamiltonian systems, driven by a weak
noise, is considered. A general method to find regions of stability in
the phase space of a randomly-driven system, based on a specific  Poincar\'e 
map, is proposed and justified. Physical manifestations of these
regions of stability, the so-called coherent clusters, are demonstrated with
two models in ocean physics. We find bunches of sound rays propagating
coherently in an underwater waveguide through a randomly fluctuating ocean at
long distances. We find clusters of passive particles to be
advected coherently by a random two-dimensional flow modelling mixing around
a topographic eddy in the ocean.
\end{abstract}
\pacs{05.45.-a, 05.40.Ca, 92.10.Vz, 47.52.+j}
\maketitle

\section{Introduction}

The interplay between dynamical chaos and noise is a topic of interest both in
nonlinear dynamics and statistical physics. Real-world systems operate 
against a noisy background. Both chaos and noise mean random or 
diffusive-like behavior in the nonlinear system`s dynamics that can be
quantified by the maximal Lyapunov exponents
\begin{equation}
\lambda=\lim_{t\to\infty}\lambda(t),\qquad
\lambda(t)=\lim_{\Delta(0)\to 0}
\frac{1}{t}\ln\frac{\Delta(t)}{\Delta(0)},
\label{0}
\end{equation}
where $\Delta(t)$ is a distance (in the Euclidean sense) in a 
direction 
at the moment $t$ between two initially closed trajectories.

However, the methods and approaches to study deterministic and random system
are different. Typical deterministic nonlinear systems (the so-called
nonhyperbolic system) are known to have mixed phase space with coexisting
regions of stable and unstable motion. There are islands of regular motion
merged into a chaotic sea, islands around islands, stochastic layers confined
between invariant tori, cantory nearby the boundaries of islands, and so on
(for a recent review of chaos in Hamiltonian systems see \cite {Z04}). The motion
is predictable in some regions but permits a statistical description only in
those ones where extremal sensitivity to initial conditions takes place. The
notion of the mixing time, the reciprocal maximal Lyapunov exponent, can be
introduced to quantity the so-called predictability horizon.

To describe randomly-driven systems one uses some ergodic postulates based
on the assumption of fully developed chaos.
In Hamiltonian nonintegrable systems intermittent-like dynamics with chaotic
oscillations interrupted by regular ones seems to be more realistic. It means,
in particular, that regular motion does not cease suddenly, and completely
stochastic motion does not arise suddenly just beyond the predictability horizon.
Remnants of stability should persist for more long times. It is especially
true for oscillatory-like systems due to resonant interaction of unperturbed
motion with random perturbation \cite{MUP04}. 

In this paper we treat a model of a randomly-driven nonlinear oscillator
from purely deterministic point of view, and show that there are regions of
stability surviving for a comparatively long time under a weak noisy excitation
with arbitrary spectrum. This time can be estimated with the help of 
a specific map which plays the role of the Poincar{\'e} map in 
periodically-driven systems. The map is intended to find numerically 
clusters of trajectories
with close initial conditions which are stable by Lyapunov with a given
realization of the random perturbation (\setcounter{abcd}{2}Sec.~\Roman{abcd}). We check the effectiveness
of the procedure proposed with two problems of ocean physics where, as in
other geophysical problems, there is, as a rule, no respective statistical
ensemble of averaging, and we deal with single realizations in the field
experiments. Noise-induced coherent clusters are demonstated with the ray
model of sound propagation in an ocean waveguide with a randomly fluctuating
sound-speed profile induced by internal waves in the deep ocean 
(\setcounter{abcd}{3}Sec.~\Roman{abcd}).
Another example is clustering of passive particles advected by a random 
two-dimensional velocity field in an ideal fluid 
(\setcounter{abcd}{4}Sec.~\Roman{abcd}).

\section{Effective Poincar\'e  map}

In this section we introduce an effective Poincar\'e  map that,
like to the usual Poincar\'e map,
enables to find numerically regions of stability in the phase space
of a Hamiltonian system surviving under a weak random perturbation
due to nonlinear resonances between the unperturbed motion and the perturbation.

Let us consider a
one-dimensional nonlinear oscillator with the Hamoltonian
\begin{equation}
H=H_0+\varepsilon H_1(t)=
\frac{p^2}{2}+U(q)+\varepsilon V(q)\xi(t),
\label{1}
\end{equation}
where $q$ and $p$ are position and momentum,
U(q) is an unperturbed potential,
$V(q)$ is a smooth function,
$\xi(t)$ is a weak noise with $\varepsilon\ll 1$, and 
normalized first and second moments,
$\left<\xi\right>=0$ and $\left<\xi^2\right>=1/2$.
Hereafter, we will consider a single realization
of noise, $\xi(t)$, and, therefore, the equations of motion
\begin{equation}
\frac{dq}{dt}=p,\quad
\frac{dp}{dt}=-\frac{dU}{dq}-\varepsilon\frac{dV}{dq}\xi(t),
\label{2}
\end{equation}
can be treated as deterministic ones. Introducing the canonical
transformation from the $(q,p)$ variables to the action--angle variables
(see \cite{Z04,LL} )
\begin{equation}
I=\frac{1}{2\pi}\oint p\,dq,\quad
\vartheta=\frac{\partial}{\partial I}\int\limits_{q'}^q p(q,\,H)\,dq,
\label{3}
\end{equation}
we rewrite Eqs.(\ref{2}) in the form 
\begin{equation}
\begin{aligned}
\frac{dI}{dt}=-\frac{\partial H(I,\,\vartheta)}{\partial\vartheta}=
-\varepsilon\frac{\partial V(I,\,\vartheta)}{\partial\vartheta}\xi(t), \\
\frac{d\vartheta}{dt}=\frac{\partial H(I,\,\vartheta)}{\partial I}=
\omega(I)+\varepsilon\frac{\partial V(I,\,\vartheta)}{\partial I}\xi(t),
\end{aligned}
\label{4}
\end{equation}
where $\omega(I)\equiv\partial H_0/\partial I$
is an action-dependent characteristic frequency of oscillations. 

The notion of stability means a weak sensitivity of a solution to small changes
in initial conditions that is quantified by the Lyapunov exponent 
(\ref{0}).
As we know from KAM-theory \cite{AKN}, most of the invariant tori
of integrable Hamiltonian systems, where the motion is stable,
are preserved under a small perturbation. In spite of each single
realization of a random perturbation $\xi(t)$ can be treated as a 
deterministic function than a stochastic one, infinite number
of frequencies in the spectrum of $\xi(t)$ leads to densely 
distributed resonances in the phase space and laking of invariant tori.
In the limit $t\to\infty$,
no regions in the phase space remain forbidden
under a noisy perturbation. Even a weak noise can accelerate
phase-space transport 
by forcing trajectories to traverse KAM tori
\cite{C79,RRW81,LL}.
So deterministic description ceases to have a sense, and one 
forces to use statistical description of the motion.

However, in practics, we deal with a finite time interval $[0:T_0]$.
Moreover, in geophysical field experiments experimentalists
work mainly with single realizations of nonstationary processes
of interest.
By these reasons we can recover ``deterministic'' methods to explore
the sets of stable trajectories satisfying to the condition of the
finite-time invariance:
{\it if any set in the phase space at $t=0$ transforms to itself
at $t=T_0$ without mixing, then
it corresponds to an ensemble of  
trajectories which are stable by
Lyapunov within the interval $[0:T_0]$}.
In order to find such stable sets for an arbitrary spectrum of perturbation,
we propose the following  map
\begin{equation}
\begin{array}{c}
I_{i+1}=I(t=T_0,\,I_i,\vartheta_i),\quad
\vartheta_{i+1}=\vartheta(t=T_0,\,I_i,\vartheta_i),
\end{array}
\label{5}
\end{equation}
where $I(t;\,I_i,\vartheta_i)$ and $\vartheta(t;\,I_i,\vartheta_i)$ are
the solutions of Eqs.(\ref{4}) with initial conditions
$I(0)=I_i$, $\vartheta(0)=\vartheta_i$. In fact, this map is equivalent
to a Poincar\'e map for a system with the Hamiltonian 
\begin{equation}
H=H_0(I)+V(I,\,\vartheta)\tilde\xi(t),
\label{6}
\end{equation}
where $\tilde\xi(t)$ is a periodic function consisting 
of identical pieces of $\xi(t)$ of the same duration $T_0$
\begin{equation}
\tilde\xi(t+nT_0)=\xi(t),\quad
0\le t\le T_0,\quad n=0,1, \dots .
\label{7}
\end{equation}
In this way we replace our original randomly-driven system by a
periodically-driven one. It should be emphasized that the validity 
of this replacement in restricted by the time interval $[0:T_0]$.
By analogy with the usual Poincar\'e map, the key property of the 
map (\ref{5}) is the following:
{\it each point of a continuous closed trajectory of the 
map (\ref{5}) corresponds to a starting point
of the solution of Eqs.(\ref{4}) 
which remains stable by Lyapunov till the time $T_0$}.
The inverse statement is not, in general, true.
It will be shown below that the map (\ref{5}) provides sufficient
but not necessary criterion of stability.
Topological properties of trajectories of the map (\ref{5})
can be treated in the framework of the theory of nonlinear resonance
\cite{C79}.
The functions $V(I,\,\vartheta)$ and $\tilde\xi(t)$ can be decomposed
in Fourier series 
\begin{equation}
\begin{aligned}
V(I,\,\vartheta)=
\sum\limits_{l=-\infty}^{\infty}V_l(I)\,\exp[i(l\vartheta+\phi_l)],\\
\tilde\xi(t)=\sum\limits_{m=-\infty}^{\infty}
\xi_{m}\,\exp[i(m\Omega t+\psi_m)],
\end{aligned}
\label{8}
\end{equation}
where $\Omega=2\pi/T_0$. The Fourier amplitudes of the analytic
function $V$ decay as $V_l(I)\sim\exp(-\sigma l)$, whereas 
the ones for the random function $\xi(t)$ decay as
$\xi_m\sim m^{-\beta}$  ($0 \le  \beta \le 1$ and  $\beta=0$ for a white 
noise) since $\xi(0)\ne\xi(T_0)$.
If the interval $T_0$ is large enough, the parameter $\beta$
is determined mainly by the correlation time of $\xi(t)$.

Substituting the expansions (\ref{8}) into the equations of motion
(\ref{4}) and omitting rapidly oscillating terms
$\exp[i(l\vartheta+m\Omega t)]$, we get the following equations:
\begin{equation}
\begin{aligned}
\frac{dI}{dt}=-\frac{i\varepsilon}{2}
\sum\limits_{l,m=0}^{\infty}
lV_l\xi_m
\exp{i\Phi}+\text{c.\ c.}, \\
\frac{d\vartheta}{dt}=\omega+\frac{\varepsilon}{2}
\sum\limits_{l,m=0}^{\infty}
\frac{\partial V_l}{\partial I}
\xi_m\exp{i\Phi}+\text{c.\ c.},
\end{aligned}
\label{9}
\end{equation}
where $\Phi=l\vartheta-m\Omega t+\phi_l-\psi_m$.
The stationary phase condition $d\Phi/dt=0$ implies 
the resonances of the map~(\ref{5})
\begin{equation}
mT(I)=lT_0,
\label{10}
\end{equation}
where $T(I)=2\pi/\omega(I)$ is a period of unperturbed oscillations
with a given value of the action $I$.
Resonant values of the action $I_\text{res}$ can be found 
from the condition (\ref{10}). The relation $l:m$ defines
the order of the respective resonance. It should be noted that an infinite
number of resonances $jl:jm$ ($j=1,2,...$) corresponds simultaneously
to each value of the resonant action. However, if $I_\text{res}$
is far enough from the separatrix value, the product $V_l\xi_m$
decreases rapidly with increasing $j$, and the resonances with
small values of $l$ and $m$ only can affect significantly a trajectory.
Thus, if $T_0>T(I_\text{res})$, only the superior term with $l=1$
should be taken into account. Neglecting the higher-order resonances,
we can describe the motion in the vicinity of $I_\text{res}$ in the
pendulum approximation \cite{C79}.
Leaving the resonant term only, we can rewrite Eqs.(\ref{9}) in the form
\begin{equation}
\begin{aligned}
\frac{dI}{dt}=\varepsilon
V_1(I_\mathrm{res})\xi_m\sin\Phi, \\
\frac{d\Phi}{dt}=\omega-m\Omega+\varepsilon
\frac{\partial V_1(I_\mathrm{res})}{\partial I}
\xi_m\cos\Phi,
\label{11}
\end{aligned}
\end{equation}
which corresponds with some simplification to the universal 
Hamiltonian of nonlinear resonance \cite{C79,Z04}
\begin{equation}
H_u=\frac{1}{2}\,\bigl|\omega'_I\bigr|\left(\Delta I\right)^2
+\varepsilon V_{l}\xi_m\cos{\Phi},
\label{12}
\end{equation}
where $\Delta I=I-I_\text{res}$.
Solutions of Eqs.(\ref{11}) describe oscilations of the action $I$ and 
phase $\Phi$ variables nearby elliptic fixed points of the 
respective resonances, the so-called phase oscillations.

An angular location of the fixed points depends on a random phase $\psi_m$ and, therefore, varies from one 
realization of the random function $\xi(t)$ to another.
A trajectory of the map (\ref{5}), being captured into a resonance, draws a chain-like structure in the phase space.
In the regime of stable motion neighboring chains are far enough from each 
other, and the space between them
is filled by non-resonant stable trajectories. The width of the resonance in terms of the frequency can be estimated
from the Hamiltonian (\ref{12}) as follows:
\begin{equation}
\Delta\omega=|\omega'_I|\Delta I_\text{max}\simeq
2\sqrt{\varepsilon|\omega'_I|V_1\xi_m},
\label{12a}
\end{equation}
where $\omega'_I=d\omega/dI$ is related to nonlinearity
of oscillations and determined by the form of the unperturbed potential.
The distance between the neighboring resonances of the successive orders
is $\delta\omega(T_0)=2\pi/T_0$. If the criterion of Chirikov, 
$\Delta\omega/\delta\omega \simeq 1$,
holds, the resonances overlap in the phase space, and the phase oscillations become
unstable.

Except for some specific situations, the Chirikov's criterion
provides a sufficient condition for emergence of global chaos.
We can estimate from the above relations the time that is required
for overlapping all the resonances in the phase space
\begin{equation}
T_\text{c}\simeq\frac{\pi}
{\sqrt{\varepsilon|\omega'_I|V_1(I_\text{min})\xi_m}},
\label{13}
\end{equation}
where $I_\text{min}$ is a minimal resonant action corresponding 
to the deepest resonance, and the amplitudes $\xi_m$ are supposed to be 
independent on time. Nevertheless, some stable regions
in the phase space can survive at $T>T_\text{c}$ in the form of
islands submerged into a stochastic sea. The time horizon $T_\text{c}$
may be treated as a time of mixing in the phase space. With a given 
spectrum of noise and large enough values of $T_0$, difference between 
the amplitudes $\xi_m$ for different realizations of the noise becomes  
negligible. Therefore, the time of phase correlations $T_\text{c}$ 
is independent on the realization chosen and may be treated as a 
characteristic quantity of a Hamiltonian system under consideration.

\section{Clustering of sound rays in an ocean waveguide}

Low-frequency acoustic  signals may propagate in the deep ocean to very long
ranges due to existence of the underwater sound channel which acts as a
waveguide confining sound waves within a restricted water volume (see general
theory in Ref.~\cite{BL91} and description of field experiments in the Pacific
ocean in the issue \cite{JASA}). In the geometric optics approximation, the
underwater sound ray trajectories in a two-dimensional waveguide in the deep
ocean with the sound speed $c$, being a smooth function of depth $z$ and range
$r$, satisfy the canonical Hamiltonian equations \cite{LaL}
\begin{equation}
\frac{dz}{dr}=\frac{\partial H}{\partial p}, \qquad
\frac{dp}{dr}=-\frac{\partial H}{\partial z},
\label{ray_eq1}
\end{equation}
with the Hamiltonian
\begin{equation}
H=-\sqrt{n^2(z,r)-p^2},
\label{ham-init}
\end{equation}
where $n(z,r)=c_0/c(z,r)$ is the refractive index, $c_0$ is a reference sound
speed, $p=n\sin\phi$ is the analog to mechanical momentum, and $\phi$ is a ray
grazing angle. In the paraxial approximation (that takes into account rays
launched at comparatively small grazing angles that can propagate in the ocean
at long distance without touching the lossy bottom), the Hamiltonian can be
written in a simple form as a sum of the range-independent and range-dependent
parts
\begin{equation}
H=H_0+H_1(r)
\label{acoust_ham01}
\end{equation}
with 
\begin{equation}
H_0=-1+\frac{p^2}{2}+\frac{\Delta c(z)}{c_0},\quad
H_1(r)=\frac{\delta c(z,r)}{c_0},
\label{acoust_ham}
\end{equation}
where $\Delta c(z)=c(z)-c_0$ describes variations of the sound speed along the
waveguide and $p\simeq \tan\phi$. The term $H_1$ describes variations of the
sound speed caused, mainly, by internal waves in ocean.

In a simplified model of perturbation with $\delta c$ being a periodic function
of the range $r$, extremal sensitivity of ray trajectories to initial
conditions~--- ray dynamical chaos~--- has been found \cite{AZ81,AZ91}. In a
number of recent publications \cite{SFW,BCT,WT01,MUP04}, it has been realized
that ray chaos should play an important role in interpretating the long-range
field experiments \cite{D92,W99,JASA}. More realistic models should include not
only horizontal but vertical modes of internal waves as well. Moreover, the
internal waves perturbation is stochastic in the real ocean.

We use a background sound-speed profile introduced in Ref.~\cite{MUP03} to
model sound propagation through the deep ocean
\begin{equation}
c(z)=
c_0\left(1-\frac{b^2}{2}(\mu-e^{-az})(e^{-az}-\gamma)\right),\quad
0\le z\le h,
\label{prof-s}
\end{equation}
where $\gamma=e^{-ah}$, $\mu$, $a$, and $b$ are adjusting parameters, $h$ is the
maximal depth, and $c_0=c(h)$. We represent the perturbation of the background
sound-speed profile along the waveguide in the following form:
\begin{equation}
\frac{\delta c(z,r)}{c_0}=\varepsilon V(z)\xi(z,r),
\label{pert-a2}
\end{equation}
where $\varepsilon\ll 1$, $V(z)$ describes sound-speed fluctuations along the
vertical coordinate $z$, and $\xi(z,r)$ is a fluctuation field with
$\left<\xi\right>=0$ and $\left<\xi^2\right>=1/2$. Let us assume that the
stochastic perturbation can be written in the factorized form
\begin{equation}
\xi(z,r)=X(z)Y(r).
\label{fact-xi-acoust}
\end{equation}
Our aim is to find coherent clusters of sound rays that are stable up to a
distance $d$ from a source. After making the canonical transformation from a 
ray variables $(z,p)$ to the action-angle ones $(I,\vartheta)$, we use the
effective  Poincar\'e map (\ref{5}) which now takes the form
\begin{equation}
\begin{aligned}
&I_{i+1}=I(d;\,I_i,\vartheta_i), \\
&\vartheta_{i+1}=\vartheta(d;\,I_i,\vartheta_i).
\end{aligned}
\label{map2}
\end{equation}
Let us expand $Y(r)$ in the Fourier series
\begin{equation}
Y(r)=\frac{1}{2}\sum\limits_{m=-\infty}^{\infty}
Y_{m}\,e^{imk_rr},
\label{fourier-xi-r2}
\end{equation}
where $k_r=2\pi/d$, and represent $X(z)$ as a sum of vertical modes
\begin{equation}
X(z)=\frac{1}{2}\sum\limits_{n=-\infty}^{\infty}
X_{n}\,e^{in\pi\chi(z)},
\label{fourier-xi-z2}
\end{equation}
where $\chi(z)=e^{-z/B}-e^{-h/B}$, and $B$ is a thermocline depth. Substituting
(\ref{fourier-xi-r2}) and (\ref{fourier-xi-z2}) into the equations of motion
\begin{equation}
\frac{dI}{dr}=-\frac{1}{c_0}\frac{\partial(\delta c)}{\partial\vartheta},\qquad
\frac{d\vartheta}{dr}=\omega+\frac{1}{c_0}\frac{\partial(\delta c)}{\partial I},
\label{req}
\end{equation}
one gets the resonance conditions
\begin{equation}
\begin{aligned}
&l\omega-mk_r\pm n\pi p\frac{e^{-z/B}}{B}=0,\\
&l\omega+mk_r\pm n\pi p\frac{e^{-z/B}}{B}=0,
\end{aligned}
\label{rescond}
\end{equation}
where $l$, $m$, and $n$ take non-negative values only.

As it is seen from (\ref{rescond}), vertical oscillations of the speed of sound
give an additional term in the resonance conditions proportional to $p$ and
produce a special type of nonlinear resonances which can break regions of
stability in the phase space \cite{M04,U05}. So the question is: could coherent
ray clusters survive under a combined horizontal and vertical stochastic
perturbation? The horizontal perturbation  $Y(r)$ is modelled as a sum of a
large (of the order of one thousand) number of harmonics with random phases and
the wave numbers distributed  in the range from
$k_\text{min}=2\pi/100$ km${}^{-1}$ to $k_\text{max}=2\pi$ km${}^{-1}$. The
spectral density of the perturbation decays as $k^{-2}$. For more information
about modelling noise see Ref.~\cite{MUP04}.

\begin{figure}[!htb]
\begin{center}
\includegraphics[width=0.45\textwidth]{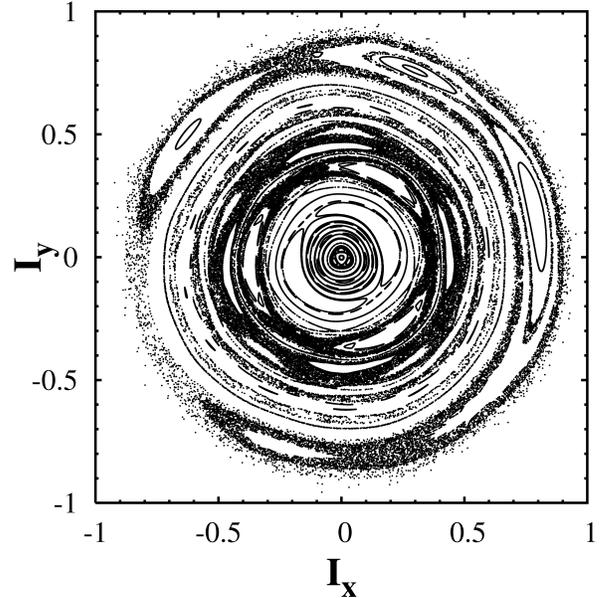}
\caption{The map (\ref{map2}) with $d=100$~km revealing coherent
ray clusters surviving in a model underwater waveguide under a
noisy-like internal-wave perturbation (\ref{fact-xi-acoust})
with a single vertical mode (\ref{x-z1}). The polar action-angle variables 
$I_x=(I/I_s)\cos \theta$ and $I_y=(I/I_s)\sin \theta$ are in units of 
the separatrix action $I_s$.}
\label{Fig1}
\end{center}
\end{figure}
\begin{figure}[!htb]
\begin{center}
\includegraphics[width=0.45\textwidth]{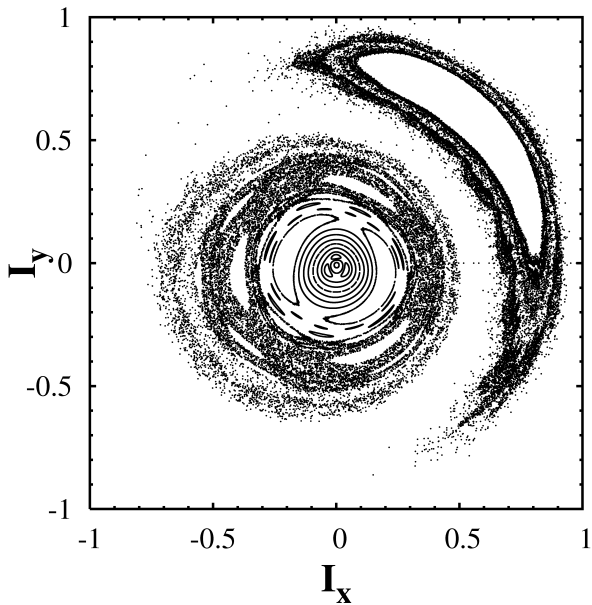}
\caption{The same as in Fig.~1 but with five vertical modes (\ref{x-z5}), 
$d=100$~km.}
\label{Fig2}
\end{center}
\end{figure}
Let the vertical oscillations contain a single mode only
\begin{equation}
X(z)=X_1\sin{\pi\chi(z)},
\label{x-z1}
\end{equation}
and the other parameters are as follows: $\varepsilon=0.0025$,
$V(z)=(2z/B)e^{-2z/B}$, and $B=1$~km. We start with a large number of 
initial values of the ray action-angle variables in different regions 
of the phase space $(I_0, \theta_0)$ and compute their values at  
the distance $d=100$~km from a sound source with the respective equations 
of ray motion   with a chosen realization of the noisy-like horizontal 
internal-wave perturbation (\ref{pert-a2}) and the single vertical 
mode (\ref{x-z1}). Then we compute the ray variables with the same realization 
of noise at the same distance $d=100$~km, find $(I_1, \theta_1)$, and so on. 
The result of computing of the Poincar\'e 
map in the coordinates $I_x=(I/I_s)\cos\vartheta$ and
$I_y=(I/I_s)\sin\vartheta$ with  $I_s=I(H=-1)$, being the separatrix value of the 
action, is shown in
Fig.~\ref{Fig1}. The map reveals numerous island-like regions of stability
in the phase space corresponding to the resonance condition $mD(I)\simeq ld$,
where $D(I)$ is a cycle length of a ray with a given action $I$. It should be 
noted that the outer chain with four islands corresponds to the same 1:1 
resonance. Under a
multi-frequency excitation, the system, being captured into the resonance
$mD=ld$, is, simultaneously, in all the resonances $jmD=jld$, where $j$ is an
integer. 
\begin{figure}[!htb]
\begin{center}
\includegraphics[width=0.45\textwidth]{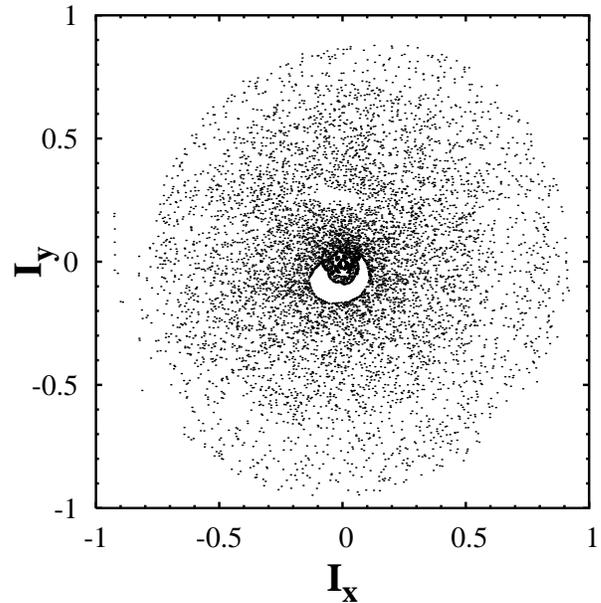}
\caption{The same as in Fig.~2 but with $d=500$~km.}
\label{Fig3}
\end{center}
\end{figure}
Let us consider now a more complicated model with five vertical modes of
internal waves
\begin{equation}
X(z)=\sum_{n=1}^5 X_n\sin(n\pi\chi(z)+\phi_n),
\label{x-z5}
\end{equation}
where $X_n\propto n^{-2}$, and $\phi_n$ are random phases. The respective
map (\ref{map2}) is shown in Fig.~\ref{Fig2} with $d=100$~km.
It follows from comparing Figs.~\ref{Fig1} and~\ref{Fig2} that
complication in the vertical perturbation effects, mainly, the most steep rays
corresponding to the outer resonance islands. The most outer island in
Fig.~\ref{Fig2} is separated from the other domains of stability by a layer 
of escaping rays (by escaping rays we mean the rays quiting by different
reasons the sound waveguide during their propagation). 
With increasing $d$, the area of islands decreases but
distinct islands of stability remain visible in the phase space at a rather 
long distance $d=500$~km (Fig.~\ref{Fig3}) even with five vertical modes 
of internal waves to be included in the model.

\begin{figure}[!htb]
\begin{center}
\includegraphics[width=0.45\textwidth]{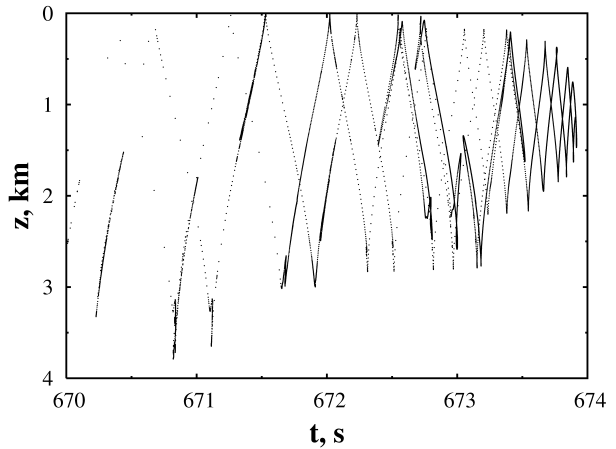}
\caption{Timefront at the range $1000$~km under the stochastic
perturbation (\ref{fourier-xi-r2}) and (\ref{x-z5}): ray depth $z$ vs ray travel time t.
Strips indicate coherent ray clusters.
}
\label{Fig4}
\end{center}
\end{figure}

The practical question
is: in which way the coherent ray clusters could be observed in real field
experiments? What is measured in the field experiments is time arrivals of
sound signals and the depths of their arrivals at a given range $R$ with the
help of a vertical receiving array of hydrophones \cite{D92,W99,JASA}. We
compute the so-called timefront at $R=1000$~km solving not the map (\ref{map2}) 
but the equations of motion (\ref{ray_eq1}) with the stochastic perturbation
(\ref{fourier-xi-r2}) and (\ref{x-z5}). The prominent strips in
Fig.~\ref{Fig4} belong, mainly, to coherent ray clusters which survive even  
at the distance of 1000 km. It should be
emphasized that similar strips have been found in the field experiments
\cite{D92,W99,JASA}.
\begin{figure}[!htb]
\begin{center}
\includegraphics[width=0.45\textwidth]{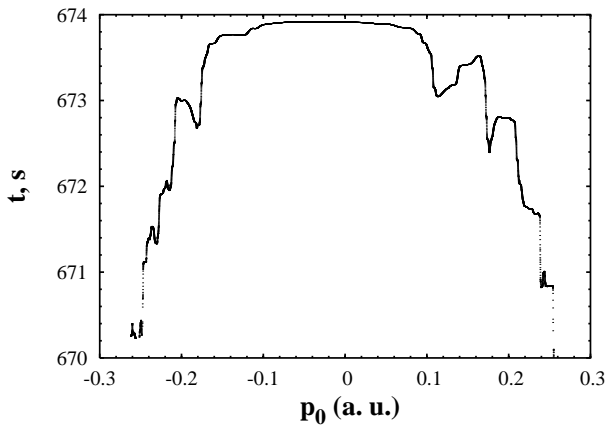}
\caption{Ray travel time $t$ vs starting momentum $p_0$ at the range $1000$~km.
Shelves indicate coherent ray clusters.
}
\label{Fig5}
\end{center}
\end{figure}

In conclusion of this section we show manifestations of coherent ray clusters in
the so-called $t-p_0$ plot. Fig.~\ref{Fig5}
demonstrates the ray travel time $t$  as a function of the 
starting ray momentum $p_0$  at the distance $R=1000$~km from 
a sound source for the model with purely horizontal random
perturbation, i.~e. Eq.(\ref{fact-xi-acoust}) with $X(z)=1$. The main
feature of this dependence is ``shelves'', i.~e. more or less flat segments in 
the $t-p_0$ plot. Each ``shelf'' corresponds to a coherent cluster. The 
``shelves'' are distributed chaotically over the range of the starting 
momenta, and their positions depend on a specific realization of the random 
perturbation. Such ``shelves'' have been found with different kinds and 
realizations of the internal-wave induced random perturbation proving 
that appearing of coherent clusters is a common feature in the sound-waveguide 
propagation through a fluctuating ocean.

\section{Clustering of passive particles in a two-dimensional flow}

In this section we consider clustering in a simple two-dimensional
flow of an ideal fluid with the dimensionless streamfunction
\begin{equation}
\displaystyle{\Psi=\ln\sqrt{x^2+y^2}+\mu x+\varepsilon x~\xi(\tau)},
\label{27}
\end{equation}
whose simplified version with $\xi(\tau)$ being a harmonic function has
been introduced in Ref.~\cite{BP01} to model advection of passive
particles in a flow with a fixed point vortex (the first term
in (\ref{27})), a stationary current along the axis $y$
with the normalized velocity $\mu$ (the second term),
and a harmonically alternating current
with the normalized velocity $\varepsilon$. In physical oceanography the
streamfunction (\ref{27}) is a simple kinematic model of mixing and transport
in the flow around a topographical eddy over a seamount in the ocean randomly
perturbed by wind in the surface layer or by small-scale turbulence.
It is well known that the Hamiltonian equations of motion for particles
(tracers) with Cartesian coordinates
$x$ and $y$ in an incompressible two-dimensional flow are written as
\begin{equation}
\begin{aligned}
\displaystyle{\dot x=-\frac{\partial \Psi}{\partial y}=
-\frac{y}{x^2+y^2}},\\
\displaystyle{\dot y=\frac{\partial \Psi}{\partial x}=
\frac{x}{x^2+y^2}+\mu+\varepsilon~\xi(\tau)}.
\end{aligned}
\label{28}
\end{equation}
\begin{figure}[!htb]
\begin{center}
\includegraphics[width=0.45\textwidth]{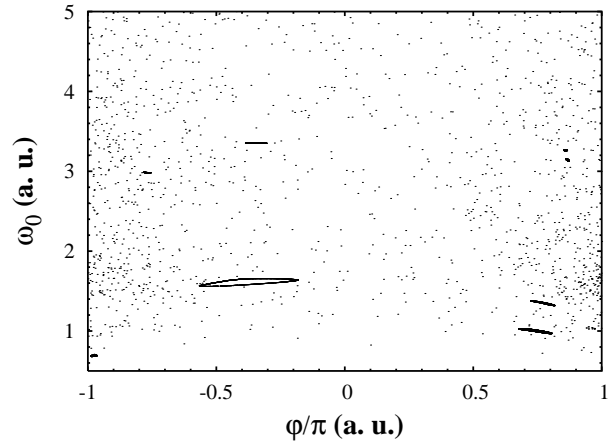}
\caption{The map (\ref{5}) with $T_0=6\pi$ revealing clusters of
advected particles in a random velocity field with the amplitude
$\varepsilon=0.01$ and the frequency band $\omega\in [2:5]$.
The unperturbed frequency of rotation of particles around
the point vortex $\omega_0$ and the polar angle $\varphi/\pi$ are in
arbitrary units.
}
\label{Fig6}
\end{center}
\end{figure}
\begin{figure}[!htb]
\begin{center}
\includegraphics[width=0.45\textwidth]{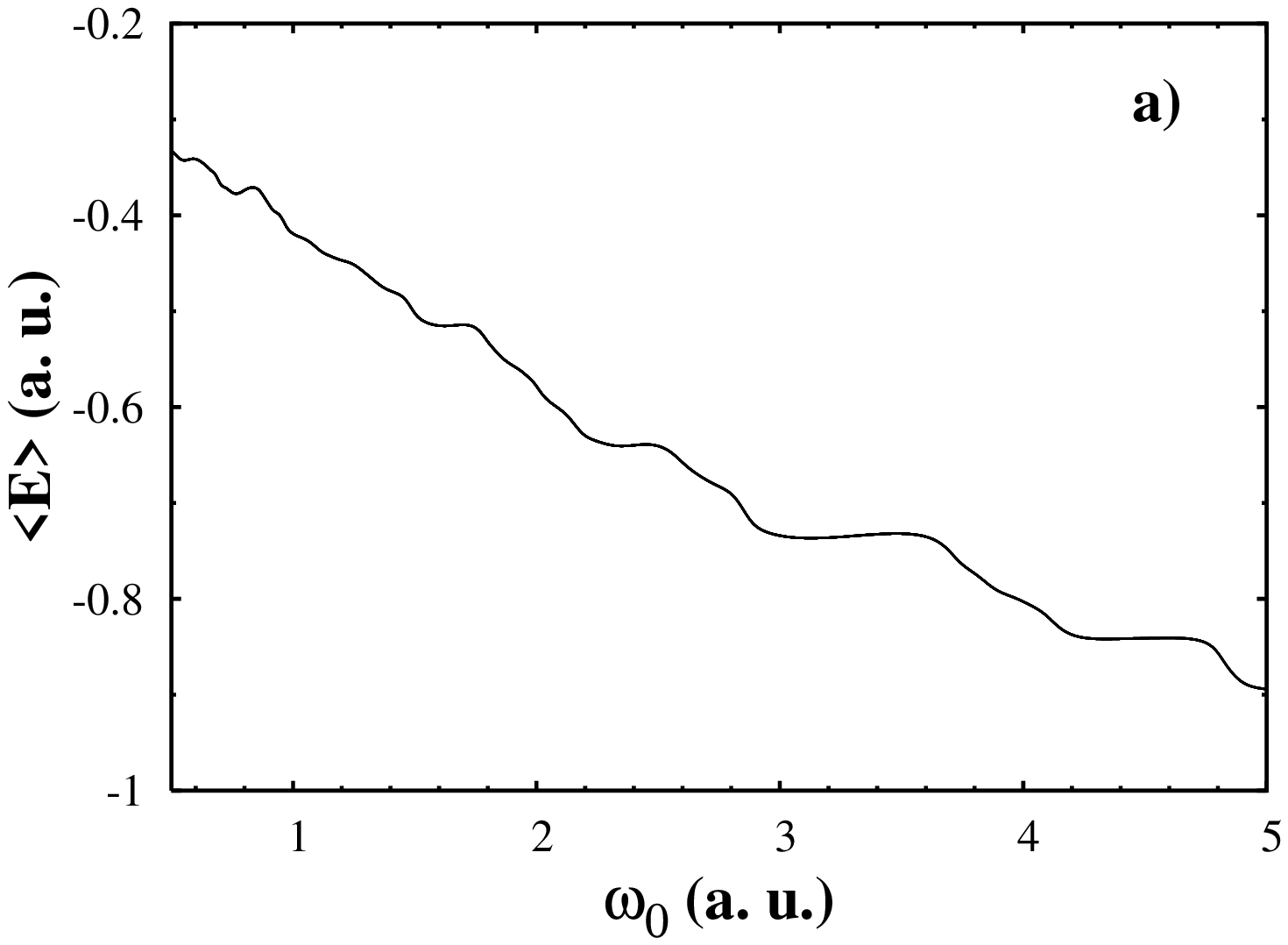}\\
\includegraphics[width=0.45\textwidth]{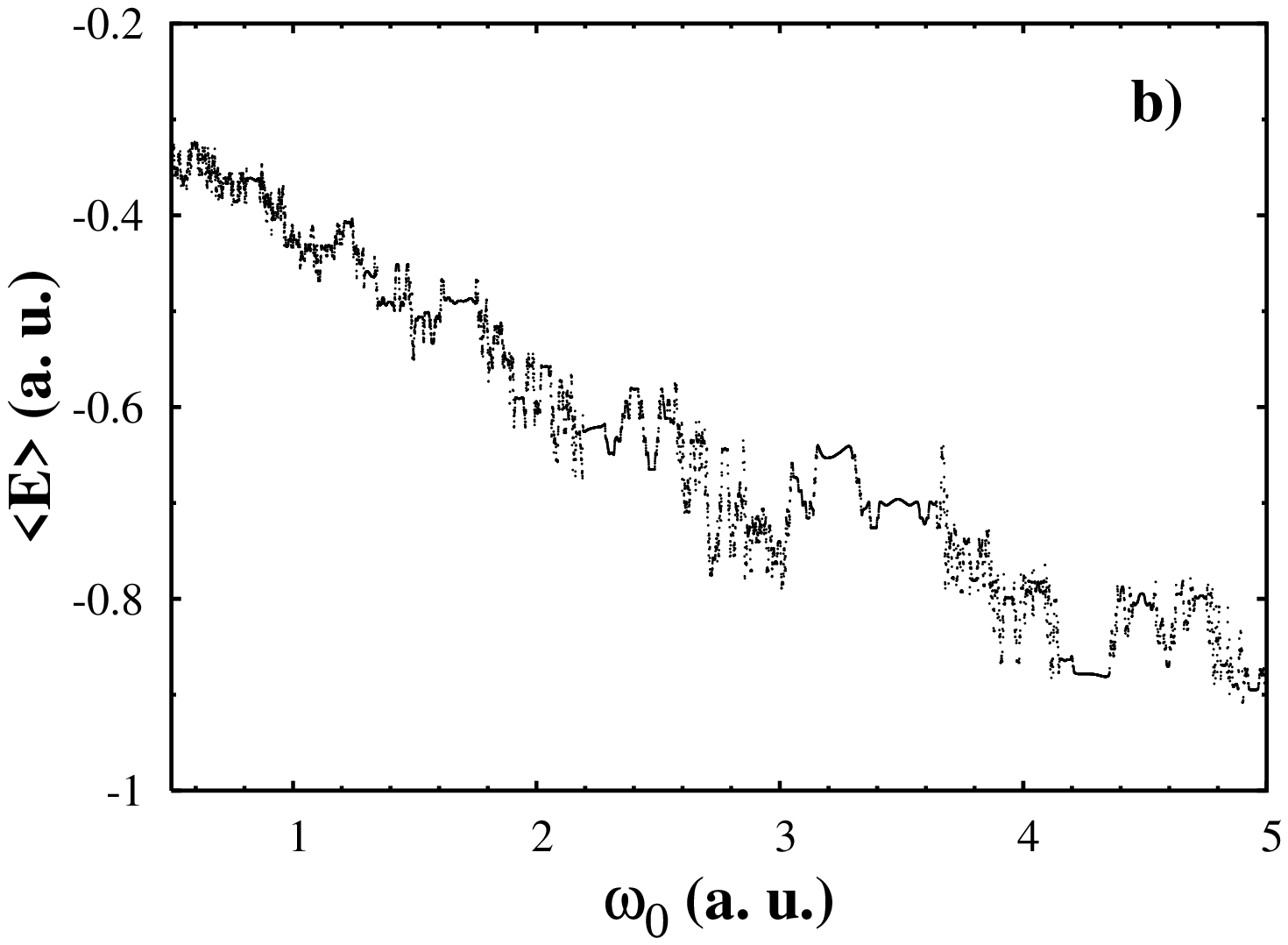}
\caption{The averaged (over $T_0$) energy of particles
$\left<E\right>$ vs their unperturbed frequency $\omega_0$:
(a) $T_0=6\pi$ and (b) $T_0=30\pi$. Shelves indicate clusters of
particles. Parameters of noise are given
in caption to Fig.~\ref{Fig6}.}
\label{Fig7}
\end{center}
\end{figure}
The configuration space of advection particles is the phase space of dynamical
system (\ref{27}), and it is possible to see noisy-induced clusters of
particles in the flow (\ref{27}) by a naked eye.
We model the noise as a random function $\xi(\tau)$ consisting of a large
number of harmonics
\begin{equation}
\displaystyle{
F(\tau)=\frac{1}{\sqrt{N+1}}\sum\limits_{n=0}^{N}\sin(\omega_n\tau+\varphi_n)},
\label{29}
\end{equation}
with random phases $\varphi_n$ and the frequencies
\begin{equation}
\displaystyle{
\omega_n=\omega_b+n\frac{\omega_e-\omega_b}{N}},\qquad
\displaystyle{n=0, 1, \dots, N},
\label{30}
\end{equation}
distributed equally in the range $[\omega_b:\omega_e]$ between
the lowest $(\omega_b)$ and highest $(\omega_e)$ frequencies.
It is possible, in principle, to model the spectrum of an arbitrary
form with the help of the series (\ref{29}) introducing an amplitude
factor depending on the frequency. Accordingly to the central limit
theorem, $\xi(\tau)$ has a Gaussian distribution with zero expectation value
because it is a large number of independent random variables.
The factor $1/\sqrt{N+1}$ provides the variance to be equal
to $1/2$.                                                

Advection of passive particles under a single-mode perturbation without
any noise has been considered it detail in Refs.\cite{PD04, JETP04}
as a scattering problem: particles enter a mixing zone around the
vortex with the incoming current and escape from it with an outcoming flow.
The scattering has been shown to be chaotic in the sense that there is a 
nonattracting invariant set, consisting of unstable periodic,
aperiodic, and marginally unstable orbits, that determines scattering and
trapping of particles from the incoming flow. The scattering has been 
shown to be fractal in the sense that scattering functions (namely, 
the dependence of the trapping time for particles and of the
number of times they wind around the vortex on the initial particle's
coordinates) are singular on a Cantor-like set of initial positions.

Now we are interested in what happens if that deterministic flow is
perturbed by a weak noise. It should be noted, first of all, 
that fractality in the scattering functions does not disappear
under noisy excitation, but the respective fractals have been
shown to have more complicated hierarchical structure
\cite{B05, U05}. In order to find noisy-induced clusters
of particles we use the effective Poincar\'e map of the type of 
(\ref{5}) with different
values of the time interval $T_0$, the amplitude of noise $\varepsilon$,
and the frequency range of noise $[\omega_b:\omega_e]$.

In Fig.~\ref{Fig6}a we demonstrate the clusters of the map of the type
(\ref{5}) with
$T_0=6\pi$ in the plane of the frequency $\omega_0$ of
rotation of particles around the vortex in the unperturbed system 
with $\varepsilon=0$ and the polar angle with $\tan \varphi=y/x$.
A large number of particles advected by the flow (\ref{27}) with 
$\mu=0.5$ and
a chosen realization of the weak noise with the amplitude $\varepsilon=0.01$ 
and the medium-frequency
range $\omega\in [2:5]$ has been chosen for simulation. 

Indirect
manifestations of clusters are seen in Fig.~\ref{Fig7}a as shelf-like segments
in the dependence of the averaged (over $T_0$) energy of tracers
$\left<E\right>$ on their unperturbed frequency $\omega_0$ at initial
positions. With increasing the interval of mapping $T_0$, noisy-induced
clusters reduce, of course, in their size, but even with $T_0=30\pi$,
which is larger than the average period of perturbation almost by two
orders of magnitude, the prominent shelves corresponding to small-size
clusters are still seen in Fig.~\ref{Fig7}b.
\begin{figure}[!htb]
\begin{center}
\includegraphics[width=0.45\textwidth]{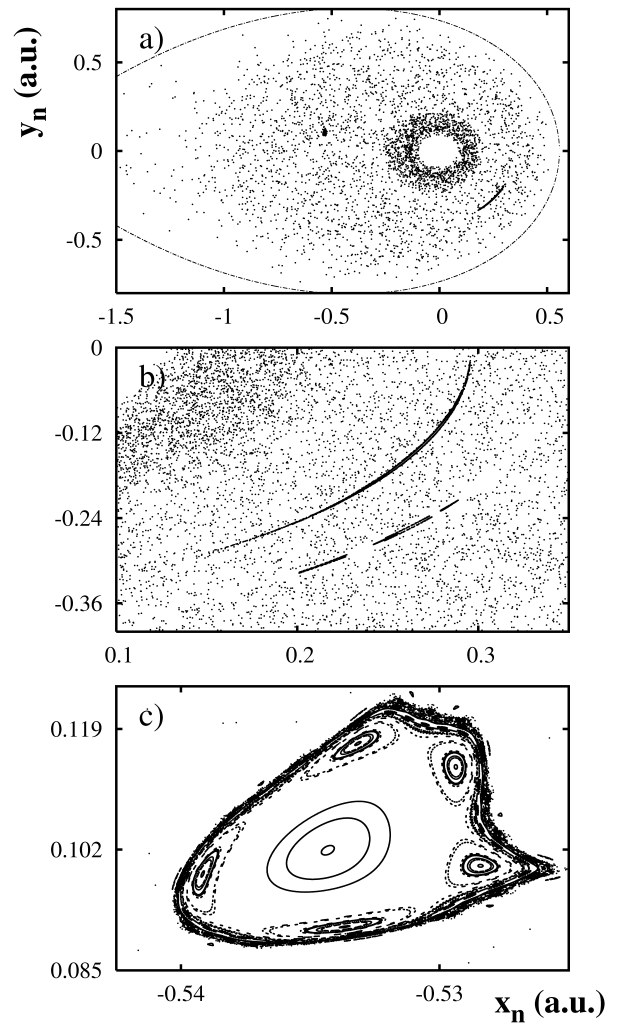}
\caption{The map (\ref{5}) with $T_0=2.3\pi$ revealing clusters
of particles in a random velocity field with the amplitude
$\varepsilon=0.1$ and the frequency band $\omega\in[9:10]$.
(a) general view in the phase plane $(x, y)$, (b) and (c)
magnifications of the respective clusters in (a). The dotted line
is an unperturbed separatrix.
}
\label{Fig8}
\end{center}
\end{figure}
\begin{figure}[!htb]
\begin{center}
\includegraphics[width=0.45\textwidth]{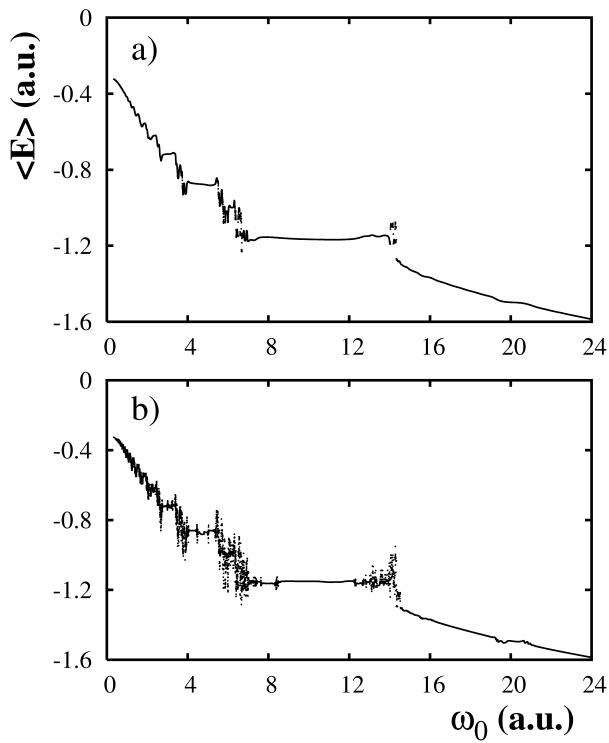}
\caption{The averaged (over $T_0$) energy $\left<E\right>$ of
particles vs their unperturbed frequency $\omega_0$:
(a) $T_0=2.3\pi$ and (b) $T_0=10\pi$. Shelves indicate clusters of
particles. Parameters of noise are given
in caption to Fig.~\ref{Fig8}.
}
\label{Fig9}
\end{center}
\end{figure}
\begin{figure*}[!htb]
\begin{center}
\includegraphics[width=0.9\textwidth]{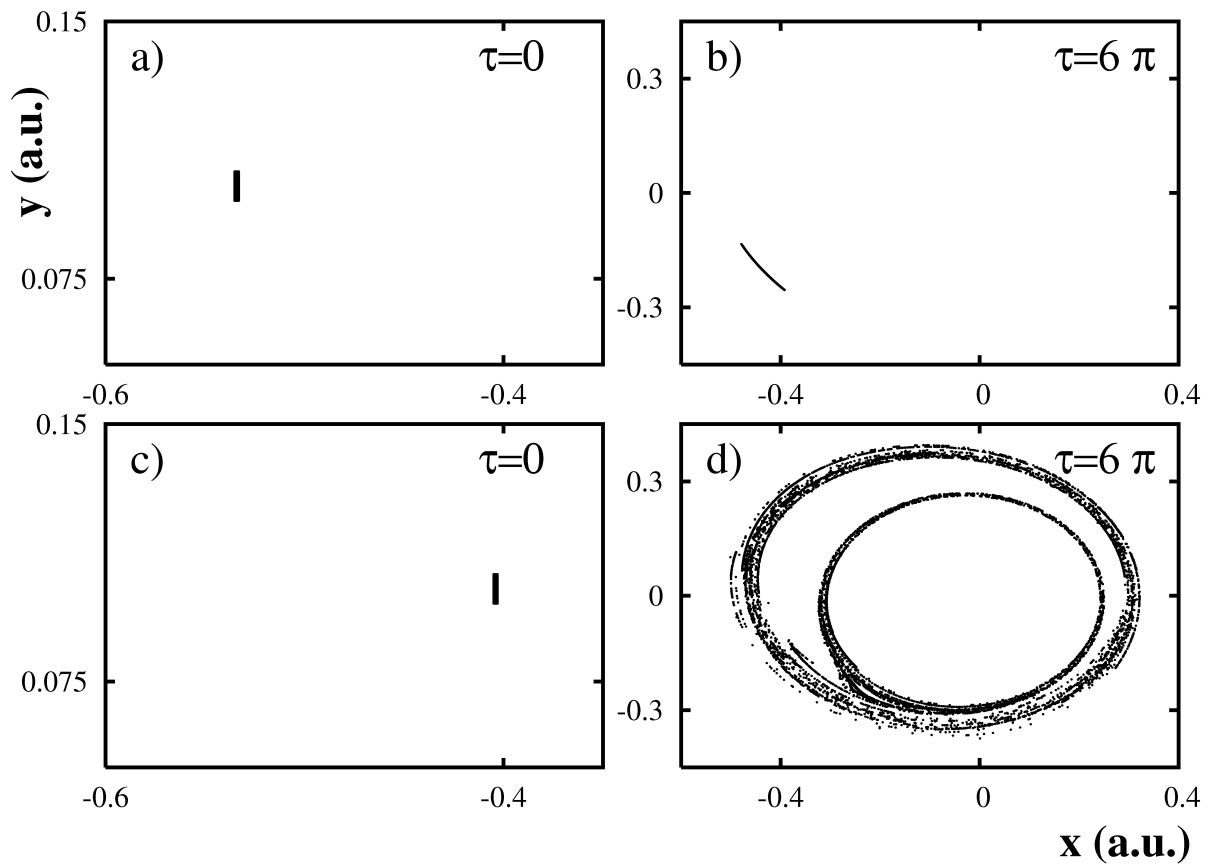}
\caption{The compact evolution of a coherent cluster of particles 
(the upper panel) and  
deformation of an unstable patch of particles (the lower panel) in a random 
velocity field. Parameters of noise are given
in caption to Fig.~\ref{Fig8}.
}
\label{Fig10}
\end{center}
\end{figure*}

In order to test the effectiveness of the map of the type 
(\ref{5}) in ``catching''
the noisy-induced clusters, we have carried out a series of numerical
experiments with $\mu=0.5$ and 
the noise of the moderate amplitude $\varepsilon=0.1$ 
and the high-frequency range $\omega\in [9:10]$. Fig.~\ref{Fig8}a demonstrates 
the result of mapping with $T_0=2.3\pi$ on the configuration plane
$(x, y)$, and Figs.~\ref{Fig8}b and c are magnifications of the respective
clusters in Fig.~\ref{Fig8}a. Prominent chains of the islands of stability 
can survive
under a rather strong noisy perturbation for a comparatively
long time. Fig.~\ref{Fig9} demonstrates shelf-like manifestations  
of noisy-induced clusters with $T_0=2.3\pi$ and  $T_0=10\pi$. 

To give a more direct manifestation of coherent clustering in a random flow 
that could be observed in real laboratory experiments with dye, we 
compare the evolution of patches of particles  chosen in the  
regions of stability and instability in the configuration space. 
In the upper panel of 
Fig.~\ref{Fig10} we show the evolution of a coherent cluster 
corresponding to the small black point (the region of stability for 
a given realization of noise) with coordinates 
$x \simeq -0.55$ and $y \simeq 0.1$ in Fig.~\ref{Fig8}a. More or less  
compact evolution of this patch with $10^4$ particles goes on up to, 
at least, $\tau = 6\pi$. For comparison, the lower panel of Fig.~\ref{Fig10}  
demonstrates on the same time interval the evolution of the patch 
with the same number of particles 
chosen initially close to the coherent cluster at 
$\tau =0$. The respective initial patch is deformed strongly at $\tau = 6\pi$. 

Clustering in the ocean is a common feature that can be seen, for
example, in satellite images of the ocean surface. In resent
years new observational tools~--- quasi-Lagrangian current following
floats and drifters~--- have been used to observe velocity
field in the ocean at different levels of depth (for a review see \cite{D91}).
In connection with our numerical observation of coherent clusters
of passive particles in the simple kinematic ocean model, we would
like to pay attention to the results of the SOFAR floats program in the
POLYMODE experiment in the North Atlantic \cite{RPW86}. Up to
forty neutrally buoyant floats at $700$~m and $1300$~m were used
to provide a quasi-Lagrangian description of the structure and evolution
of the mesoscale eddy field. Those floats can be considered as quasi-passive
tracers in a weak noise environment. A large number of the deep floats revealed
remarkably coherent motion over a two-month period.

\section{Conclusion}

We proposed and justified a rather general method to find regions of stability
in the phase space of oscillatory-like Hamiltonian systems driven by a weak
noise with an arbitrary spectrum. Physical manifestations of these regions of
stability, the so-called coherent clusters, have been demonstrated with two
models in ocean physics. We have found coherent ray clusters in the model of
underwater-waveguide sound proparagition through a randomly-fluctuating
ocean and coherent clusters of passive particles in the two-dimensional flow 
modelling kinematically advection around a topographic eddy in the ocean.

In conclusion we would like to add two remarks about the properties of the
effective Poincar\'e map (\ref{5}). Firstly, the duration of the temporal interval 
$T_0$
in constructing the map can be chosen arbitrarily. Thus, if $\xi(t)$ is a stationary
random process, the regions of stability in the phase space exist at any time
moment. Secondly, the map (\ref{5}) enables to prove definitely the existence
of some regions of stability but not all of them. Really, the map, 
constructed with any given value of $T_0$, can reveal only those stable sets
which correspond to the phase oscillations nearby the fixed points of the
map. However, there exist another regions of stability looking as chaotic
ones on the map. The topology of the map changes with varying the mapping
time $T_0$, and some regions in the phase space, which look as pseudochaotic on
the map at $T_0=T_1$, become stable at $T_0=T_2>T_1$. The total area of the 
regions of
stability, survived under a weak noise at $t=T_0$, can be estimated as 
an area of
superposition of all the stable sets detected by the map (\ref{5}) with the
mapping step varying from $T_0$ to $T_c$.

\section*{Acknowledgments} 
 
This work was supported by the Federal Program ``World Ocean" of the 
Russian Government,  
by the Program ``Mathematical Methods in 
Nonlinear Dynamics'' of the Prezidium of the Russian Academy of Sciences, 
and by the Program for Basic Research of the Far Eastern Division of 
the Russian Academy of Sciences.

\end{document}